\documentclass[12pt,preprint]{aastex}
\slugcomment{to be submitted to ApJ}

\shorttitle{V471 Tauri} 
\shortauthors{Sion et al.} 

\begin{document}
\title{
Hubble Space Telescope FUV Spectra of the Post-Common-Envelope Hyades
Binary V471 Tauri} 

\author{Edward M. Sion}   
\affil{Department of Astronomy \&  Astrophysics, 
Villanova University, 
Villanova, PA 19085} 

\author{Howard E. Bond}   
\affil{Space Telescope Science Institute, 
3700 San Martin Drive, 
Baltimore, MD 21218}  

\author{Don Lindler}   
\affil{Sigma Space Corporation, 
Greenbelt, MD} 

\author{Patrick Godon\altaffilmark{1}}   
\affil{Department of Astronomy \&  Astrophysics, 
Villanova University, 
Villanova, PA 19085} 

\author{Dayal Wickramasinghe,  Lilia Ferrario}   
\affil{Mathematical Sciences Institute, 
The Australian National University, 
ACT0200, Canberra, Australia} 

\and 

\author{Jean Dupuis}   
\affil{Space Astronomy, 
Canadian Space Agency, 
6767 Route de l'Aéroport Saint-Hubert, 
Quebec J3Y 8Y9} 

\altaffiltext{1}{Visiting at the Henry A. Rowland Department of
Physics \& Astronomy,  Johns Hopkins University, Baltimore,
MD 21218}   

\begin{abstract}

We have carried out an analysis of the HST STIS archival spectra of the 
magnetic white dwarf in the Hyades eclipsing-spectroscopic, post-common 
envelope binary V471 Tauri, time resolved on the orbit and on the X-ray 
rotational phase of the magnetic white dwarf. An HST STIS spectrum obtained 
during primary eclipse reveals a host of transition region/chromospheric
emission features including N V (1238, 1242), Si IV (1393, 1402), 
C IV (1548, 1550) and He II (1640). The spectroscopic characteristics and
emission line fluxes of the transition region/chromosphere of the very active, rapidly rotating, K2V component of V471 Tauri, are compared with the emission characteristics of fast rotating K dwarfs in young open clusters.  
We have detected a number of absorption features associated with metals accreted onto the photosphere of the magnetic white dwarf from which we  derive radial velocities. All of the absorption features are modulated on the 555s rotation period of the white dwarf with maximum line strength at rotational phase 0.0 when the primary magnetic accretion region is facing the observer. The photospheric absorption features show no clear evidence of Zeeman splitting and no evidence of a correlation between their variations in strength and orbital phase. We report clear evidence of a secondary accretion pole. We derive C and Si abundances from the Si\,{\sc iv} and C\,{\sc III} features. All other absorption lines are either interstellar or associated with a region above the white dwarf and/or with coronal mass ejection events illuminated as they pass in front of the white dwarf.  

\end{abstract} 

\section{Introduction}

V471 Tauri is a short period eclipsing binary in the Hyades star cluster, 
whose components are a dK main sequence star and a hot DAZ white dwarf. 
It is the prototype of the pre-cataclysmic binary systems (Vauclair 1972; 
Paczynski 1976), which are detached close binaries containing 
a white dwarf (WD) onto which mass transfer will be commence within a 
Hubble time (Schreiber \& G\"ansicke 2003). V471 Tau is also the prototype 
of the post-common-envelope binaries. In the common envelope (CE) 
scenario (e.g., Iben \& Livio 1993) the system originally had a much
longer orbital period than its present 12.5 hr. 
When the more massive component reached the red giant (or AGB) stage, it underwent an episode of dynamical unstable mass transfer and engulfed its 
main-sequence companion in a CE. The ensuing frictional spiral down decay of the orbit eventually led to ejection 
of the common envelope due to frictional luminosity, leaving a much
closer binary containing the core of the red giant (now the DA white dwarf) and the dK
main-sequence star. The presence of such a system in the Hyades cluster 
offers a unique opportunity, because the distance, chemical composition, and current turnoff mass of the cluster are known. 
In previous work on V471 Tauri, we determined a precise mass 
($0.85 \pm 0.09M_{\odot}$) for
the white dwarf (O'Brien et al. 2001), detected direct evidence of a coronal mass ejection
(Bond et al. 2001), confirmed the rotationally modulated magnetic accretion model for the
origin of the 9.25 minute optical/FUV/X-ray oscillations (Sion et al. 1998), detected the first photospheric
metal line due to magnetic accretion (Sion et al. 1998) and showed that
the K dwarf is oversized for its mass and the WD is far too hot for its mass compared
to other WDs in the Hyades cluster (O'Brien et al.2001). This is exactly the opposite of
expectation, since the most massive WD in the cluster should be the oldest, and 
consequently the coolest. One possible explanation is that the white dwarf is the product of a 
binary merger (O'Brien et al. 2001). Moreover, the WD component of V471 Tau exhibits variations at 
soft X-ray, EUV, and optical wavelengths at a period of 9.25 min. This variability is caused by rotational
modulation of a magnetic WD, whose polar regions are darkened in the soft X-ray and EUV
bands by accreted photospheric metals and helium (Clemens et al. 1992; Barstow et al.
1992), and brightened in the optical by UV flux redistribution. 

HST observations of V471 Tauri were first carried out with GHRS for three key science 
objectives: (1) to determine $K_1$ and the white dwarf 
mass from observations at the quadrature phases; (2) look for metals lines accreted by the 
white dwarf and; (3) studies of the K2V star 
in which the white dwarf was used as a beaming probe study the K dwarf's mass loss, chromospheric and coronal 
structures at orbital phases near the ingress
and egress of primary eclipse.  
Our previous GHRS spectra 
covered only a 35\AA\  region centered on Ly Alpha, but they revealed a photospheric 
Si\,{\sc iii} 1206 \AA\  absorption line modulated in strength on the X-ray rotational phase, such that 
at X-ray minimum (when the X-ray dark accretion pole faces the observer) the Si\,{\sc iii} absorption 
appears at maximum strength. This provided the first direct confirmation of the magnetic 
accretion model for the origin of the 9.25min X-ray/EUV/optical oscillations. 
The Si\,{\sc iii} detection (see Fig. 1 in Sion et al. 1998) also marginally 
revealed Zeeman splitting into sigma+ and sigma- components, as expected 
when looking down the magnetic field lines when the pole is seen face on. 
The observed splitting corresponds to a polar field strength of  
$\sim$350 kG. It was the first time Zeeman splitting of a metallic line has 
been seen in any magnetic white dwarf, single or binary (Sion et al. 1998).
We determined a silicon abundance of 0.1 solar (the first metal abundance determined
for any magnetic white dwarf) within the accreted Si spot, which covers about 40\% of the
visible hemisphere. Assuming that the accretion and diffusion are in equilibrium, the low
abundance implies an accretion rate 4 orders of magnitude lower than the Bondi-Hoyle
rate that would occur in the absence of a magnetic field. The highly inefficient accretion
strongly suggests the operation of a magneto-centrifugal propeller, and is the first direct 
evidence for the operation of this mechanism in any astrophysical setting. In the standard
propeller formulation, e.g. Pringle \& Rees (1972), a field of 350 kG is more than sufficient
for its operation in the case of V471 Tau. 

HST observations of V471 Tauri using STIS were conducted by several independent investigators
following the same lines of investigation as the GHRS studies mentioned above but
with greater sensitivity and much broader wavelength coverage.

The purpose of this paper is to analyze all of the subsequently obtained 
STIS spectra in the HST archive to widen and deepen our investigations. Our study of the entire STIS echelle archive, has the advantage of
higher S/N spectra at many different orbital phases and of covering a full 600\AA\ . This enables studies 
of the line variability as a function of WD rotational phase, sampling 
magnetic accretion in the face of very short diffusion timescales, 
determine chemical abundances, detect Zeeman splitting for a greater mix of ion species,
thus pin down the magnetic field strength and its variation at the rotational period. 
The time-resolved STIS spectra yield precise
radial velocities which allow clear separation between line features arising from gas in corotation
with the K dwarf, the gravitationally-redshifted photosphere of the white dwarf,
the interstellar medium and the photosphere of the K dwarf.

\section{HST STIS Observations}

An observing log of the archival STIS observations is provided in Table 1 where we list, by column,
(1) the observation number; (2) Entry Rootname; (3) Aperture; (4) Observation date; (5) The start time of the observation in UT;
(6) Exposure time in seconds; (7) the average orbital phase during the observation and; (8) comments. The STIS data included 
observations in which transient wind or coronal 
mass ejection (CME) features appear in some of the groups due 
to blobs of gas being silhouetted against the white dwarf continuum.
These features will be discussed elsewhere and observations containing 
such features were excluded from the 
present analysis. Radial velocities were computed using the average 
for data between rotational phases 0.8 to 0.2 for each of the observations tabulated below.

\begin{deluxetable}{cccccccc}
\hspace{-2.cm} 
\tablewidth{0pt}
\tablecaption{Archival HST STIS Observations}
\tablehead{
 entry       & aperture &  date    &  time     &  exp.time &  avg orbital  &   comments \\ 
         & rootname    &           & yyyy-mm-dd & hh:mm:ss  & (sec)     &    phase      &        
}
\startdata 
  o4mu02010  &  0.2X0.06 & 1998-03-13 & 04:09:07  &  1680  &  0.93 &        \\ 
  o4mua2010  &  0.2X0.06 & 1998-03-13 & 05:26:48  &  2580  &  0.06 & CMEs, Eclipse        \\ 
  o4mua2020  &  0.2X0.06 & 1998-03-13 & 07:01:39  &  2460  &  0.17 & CMEs        \\ 
  o4mu01010  &  0.2X0.06 & 1998-03-23 & 23:50:15  &  1680  &  0.69 &        \\ 
  o4mu01020  &  0.2X0.06 & 1998-03-24 & 01:06:00  &  2580  &  0.80 &        \\ 
  o4mua1010  &  0.2X0.06 & 1998-03-24 & 02:47:52  &  2370  &  0.94 &        \\ 
  o5dma1010  &  0.2X0.2  & 2000-08-24 & 17:10:35  &  1962  &  0.23 &        \\ 
  o5dma4010  &  0.2X0.2  & 2000-08-24 & 23:36:44  &  1962  &  0.74 &        \\ 
  o5dma2010  &  0.2X0.2  & 2000-08-25 & 18:43:24  &  1962  &  0.27 &        \\ 
  o5dma3010  &  0.2X0.2  & 2000-08-27 & 14:21:23  &  1962  &  0.76 &        \\ 
  o6jc01010  &  0.2X0.06 & 2002-01-25 & 01:31:52  &  1650  &  0.79 &        \\ 
  o6jc01020  &  0.2X0.06 & 2002-01-25 & 02:46:51  &  2895  &  0.90 &  CMEs        \\ 
  o6jc01030  &  0.2X0.06 & 2002-01-25 & 04:22:59  &  2895  &  0.05 &  Eclipse         \\ 
  o6jc01040  &  0.2X0.06 & 2002-01-25 & 05:59:06  &  2895  &  0.15 &        \\ 
\enddata
\end{deluxetable} 

\clearpage 

Each of the 11 STIS observations was subdivided into 16 rotational subgroups 
(phase bins), each of duration 126 s so that we have a total of 176 individual 
spectra each with a duration of $\sim$0.0625 of the white dwarf rotation period.  
After correcting the mid-exposure times to the solar system barycenter, 
the subgroups were all shifted to zero radial velocity in the rest frame 
of the WD and then co-added into four bins sorted by X-ray rotational phase.
For the white dwarf rotational phase, we used the timings of Clemens et al. (1992) to infer the rotation phase 0.0.

The velocity shift to zero radial velocity in the rest frame of the white dwarf
was carried out using the parameters computed from the SI IV 1393.755 line (Gamma = 69, $K_{wd} = 226$). The ephemeris zero point was adjusted to make the center of the large peak (optical maximum) in the rotational profile at zero phase. 

To compute the ephemeris, we divided the STIS observations (ignoring the data taken during eclipse) into 32 phase bins. We did not generate any spectra at this 
stage but only generated the light profile for the entire wavelength range.
Using the derived ephemeris, we then constructed 3062 separate images using all of the time tag data. Each image was calibrated with CALSTIS to generate 3062 spectra. These spectra were radial velocity corrected (using the fit to the SI IV 1393.755 line), sorted into four phase bins centered at phases 0, 0.25, 0.5 and 0.75 (excluding data taken during the eclipses and observations with CMEs), and averaged for each bin. The first bin was chosen to coincide with soft X-ray minimum with the other three bins equally spaced around the 555s oscillation period. In Fig.1, we display the two phases of the rotational profile, computed from the STIS data with the rotational ephemeris  HJED2450885.67019 + 0.0064193953E due to Clemens et al. (1992).
The first peak corresponds to the primary accretion pole (optical maximum, X-ray minimum) and the second peak to the clear FUV detection for the first time of the secondary accretion pole. 

\section{Eclipse Data and the K2V Component}

V471 Tauri's eclipses are characterized by a very rapid light decrease at ingress and a very steep light increase on the rise to egress. The brightness variations
due to the eclipse of the white dwarf by the K2V star have a duration only 55 seconds. In order to pinpoint the orbital phases precisely for the time of the STIS observations, eclipse timings collected by Ibanoglu et al.(2005) and covering the time span of the three STIS observations were used. By adjusting for the O-C offsets between the observations 
and the predicted eclipse times, we found that the following formulae, with the times given in Heliocentric 
Julian Ephemeris Dates (HJED) give the eclipse times to sufficient accuracy, generally +/- 20 seconds, 
at the dates of the STIS observations:

1998 Mar: HJED 2440610.06412 + 0.521183398 E 

2000 Aug: HJED 2440610.06520 + 0.521183398 E

2002 Jan: HJED 2440610.06600 + 0.521183398 E

These eclipse times are given as Heliocentric Julian Ephemeris Dates (HJED)
that differ slightly from the Heliocentric Julian Dates (HJD), which are calculated from the Universal Time. We used a program written by one of us
(HEB) in which the date/time at the beginning of each STIS observation was used as input. The transformation from HJD to HJED was 63.184 seconds for the 1998 eclipse time and 64.184 seconds for the 2000 and 2002 eclipse times, since one leap second was added to the HJED in 2000. 

In Fig. 2, we display the number of counts versus time in seconds 
obtained in eclipse egress versus orbital phase for the 1998
STIS data set where the data has been tabulated on 0.2 second 
intervals. In Fig.3, we display the eclipse egress for the 2002 
data in number of counts versus time in seconds. The data in Fig.3 has also been 
tabulated on 0.2 second intervals. The minima of both eclipses are extremely flat indicative of no pronounced surface activity on the K2V star at minimum light. 
We found that the ephemeris matched the two eclipse egress events to within 18 seconds. This accuracy is more than sufficient for our radial velocity computations. In Fig. 4, a detailed comparative overplot is given of the 1998 and 2002 eclipse data versus orbital phase. The small difference between the two light curves is probably due to a small
error in the ephemeris. Since the errors in the ephemeris are +/- 20 seconds
or 0.0005 (in phase), this is consistent with the errors in the plot.

Since the K2V star in V471 Tauri is tidally locked and rapidly rotating with Vsini = 91 km/s, it contains one of the most active known K dwarfs, and provides a nearly ideal testing lab for the theory of stellar dynamos and magnetic surface activity and structures. This is because the K2V star eclipses the white dwarf, thus providing a rare oportunity to use the white dwarf as a direct beaming probe of the K2V star's outer atmosphere, including its transition region and chromospheric/coronal structures. It is magnetically active, displaying
flares (e.g., Young et al. 1983), a photometric wave (e.g.,
Skillman \& Patterson 1988), coronal X-ray emission (e.g.,
Wheatley 1998), and emission in the chromospheric lines of
Ca II H\&K and H$\alpha$ (Skillman \& Patterson 1988). Doppler
images (Ramseyer, Hatzes, \& Jablonski 1995) reveal a large
high-latitude spot and coronal mass ejections have been detected (Bond et al. 2001; O'Brien et al.2001). The rapid rotation sustains the magnetic activity in the K dwarf at extremely high levels relative to the solar analogue. V471 Tauri's K2V star, despite being very old, has properties
similar to those of the single ultra-rapid rotators such as
AB Dor and PZ Tel (Walter 1999, 2004).

Bond et al.(2001) reported direct spectroscopic evidence of coronal mass ejections (CMEs) from the magnetically active K dwarf, larger in scale, higher in energy release and more frequently occurring than in the sun. Walter (2004) carried out intense coverage of the eclipse ingress and egress of the white dwarf using HST STIS. His study revealed very hot ($>$ 250,000K) spatially extended discrete magnetic structures around the K dwarf and a temperature inversion above the K dwarf photosphere. The magnetically confined gas was found to be in co-rotation with the K dwarf. 

We have further explored magnetic activity associated with the K dwarf. The STIS spectrum obtained during the primary eclipse is displayed in Fig. 5. It reveals a multitude of important chromospheric lines as well as H\,{\sc i} and D\,{\sc i}. These emission features include  N V (1238, 1242), He II (1640), C IV (1548,1550) and Si IV (1393, 1402), which arise from a range of high temperatures associated with different layers of the K2V chromosphere/transition region. The observed emission line strengths during primary eclipse, their wavelengths and emission line fluxes are given in Table 2.

\clearpage 

\begin{deluxetable}{ccc}
\tablewidth{0pt}
\tablecaption{Observed Emission Lines Fluxes During Primary Eclipse}  
\tablehead{
Ion      &  Wavelength     &    Flux     \\  
         &                 & ($10^{-14}$ergs/cm$^2$/s)
}
\startdata 
C III  &         1175.71    &           5.5        \\         
Si III   &        1206.53    &          3.5       \\   
N V      &        1238.82     &         2.5      \\   
N V      &        1242.80      &        2.5        \\   
Si III   &        1304.37       &       1.0        \\   
O I      &        1304.86       &       1.0        \\   
C II     &        1334.53       &       3.0        \\   
C II     &        1335.71        &      3.0        \\   
Si IV    &         1393.76       &      2.0       \\   
Si IV    &         1402.77       &      1.0       \\   
C IV      &       1548.20        &      9.0       \\   
C IV      &       1550.77         &     4.5       \\   
He II     &       1640.474        &     5.0     
\enddata
\end{deluxetable} 

\clearpage 

The strongest emission lines in Table 2 are C III (1175.71), C IV (1548, 1550), 
and He II (1640.474). It is useful to compare the emission line fluxes 
and mix of detected emission features in active single late-type 
dwarfs with chromospheric lines such as O I and C II lines associated with temperatures between 8000 and 10,000K) and the $10^5$K emission associated with the He II, N V and C IV chromospheric emission (Walter 2004). Among cool stars which are classified "hyperactive" (e.g. H II 314, two Alpha Persei stars; Ayres et al. 1996) emission lines O I (1304), C II (1335), Si IV (1400), and C IV (1549) are identified while in a number of other active G-type dwarfs, only the C IV (1550) feature is detectable. Since the cool component in V471 Tauri is classified K2V  with a rotational period of 12.5 hours (assuming synchronism with the orbital period), we compared the line detections in Table 2 with several active chromosphere K dwarfs (Simon \& Fekel 1987). For comparison, one of the closest but only moderately active K2V stars, Epsilon Eridani (Ayres et al. 1983), reveals emission lines of O I + Si III 1304, C II 1334-35, Si IV (1393, 1402) C IV 1548, 1550, He II 1640, Si II 1808, S III 1892, and C I 1994. There is no emission line due to N V (1238. 1242) in Epsilon Eri. However, the C IV 1548, 1550) emission line flux at Earth is $9.2 \times 10^{-13}$ergs/cm$^2$/s/\AA . In a chromospherically active K dwarf with a rotation period of 1.7 days, HD82558, the C IV flux observed at Earth is
$ 5.7 \times 10^{-13}$ergs/cm$^2$/s/\AA and the N V (1238, 1242) flux is $7.4 \times 10^{-14}$, giving a C IV/N V flux ratio of 7.7 whereas this same ratio in V471 Tauri is 3.6, about half as large. V775 Her, a K1 dwarf with a rotation period of 2.9 days, has a C IV/N V flux ratio of 2.5 which is closer to the measured emission line flux ratio for V471 Tauri. However, since the rotation rate of the K dwarf in V471 Tauri is roughly 0.5 days, its stellar dynamo and associated surface activity and magnetic structures may be comparable to the fastest rotating K dwarfs seen in relatively young open star clusters (e.g. NGC2547) which contain K dwarfs with rotation periods below 0.5 days. The coronal  activity of K dwarfs appears to saturate at $P_{rot} \sim  0.3$ days (Jeffries et al. 2010) and then decline again at even faster rotation rates.

\section{Detected Metallic Absorption in the STIS Spectra}

The primary motivation of this study was to probe the physics of magnetic accretion by the rotating magnetic white dwarf, detect photospheric lines, determine their abundances, the rate of magnetic accretion and the magnetic field strength of the white dwarf. Of the 13 STIS spectra listed in Table 1 above, we have selected one spectrum, o5DMA4010, taken at orbital phase 0.74 as being roughly characteristic of the others in terms of the number of lines detected. In Table 3, we present the detected absorption features due to metals and emission features due to C\,{\sc iv} and He\,{\sc ii} from which we calculate our radial velocities. By column, we have tabulated (1) the ion, (2) its rest wavelength, (3) its measured wavelength, (4) the equivalent width (A), and (5) whether feature is in emission or absorption. The strongest of the absorption lines is Si IV (1393, 1402). The other absorption lines present are Si\,{\sc iii} (1206.5), C\,{\sc ii}  1334.5323, C\,{\sc ii}  1335.7077, C\,{\sc iii}  1175.711, along with C\,{\sc iv}  1548.202 (in emission), He\,{\sc ii}  (1640.474 in emission) and an absorption line at 1501 that we tentatively identify as a blend of three transitions of Si III. The three candidate transitions are Si III (multiplet 36) at 1500.241, 1501.191 and 1501.870. However, there are two strong P III (multiplet 6) transitions with large lab intensities at 1501.551 and 1502.273. To our knowledge, the Si III multiplet 36 transitions have never before been observed in a white dwarf. Phosphorus lines are seen in CV white dwarfs (Sion et al.2001) and in wind outflows from CVs.

\begin{deluxetable}{ccccc}
\tablewidth{0pt}
\tablecaption{Observed Line Features at Rotation Phase 0.0}  
\tablehead{
Ion      &  Rest Wavelength & Obs. Wavelength & Equivalent Width  & Feature     \\  
         &   (\AA )        &  (\AA )              &  (\AA )       &   
}
\startdata 
C\,{\sc iii}    &   1175.711     &     1176.401   &  0.295  &   absorption    \\   
Si\,{\sc iii}   &   1206.533     &     1207.602   &  0.134  &   absorption     \\             
C\,{\sc ii}     &   1334.5323    &                &         &   absorption   \\
C\,{\sc ii}     &   1335.7077    &     1336.499   &  0.135  &   absorption      \\                                  
Si\,{\sc iv}    &   1393.755     &     1395.123   &  0.243  &   absorption       \\                           
                &   1402.140     &     1404.231   &  0.174  &   absorption  \\
Si\,{\sc iii}   &   1501.3       &                &         &   absorption  \\
C\,{\sc iv}     &   1548.202,    &      1548.653  & -0.118  &    emission  \\
                &   1550.774     &      1550.652  & -0.087  &    emission      \\                       
He\,{\sc ii}    &   1640.474     &      1640.643  & -0.25   &    emission  \\
\enddata
\end{deluxetable} 

\clearpage 

Our calculated radial velocities have typical uncertainties of 2 to 3 \%.
We found that the Si\,{\sc iv} (1393, 1402) doublet clearly tracks with the orbital motion of the white dwarf. Since the Si\,{\sc iv} 1393 component was a much cleaner line than the 1402 component, we used it to compute the radial velocity. The velocity derived from the Si\,{\sc iv} 1402 feature agrees to within 1.5 sigma. In Fig.6, we display the radial velocity curve derived for the Si\,{\sc iv} 1393.755 feature. 

In Fig.7, we display the radial velocities versus orbital phase for five absorption lines (C III (1175), C IV (1548, 1550), and Si IV 1393, 1402)) relative to the radial velocity function derived for the Si IV 1393 feature. We used the data between rotational phases 0.8 to 1.2 when the features are strongest because the primary accretion area is facing the observor. When we examined the stacked spectra, corrected for radial velocity using the Si IV 1393 line fit, the weak SiV 1402 member of the doublet agrees with the Si IV 1393 component as expected, and the C III (1175) absorption line also matches the radial velocity of the S IV 1393 line. However, it was difficult to obtain accurate radial velocities for O I + Si III (1302) and C II (1335) due to their weakness and contamination by an interstellar line. Nevertheless, there is some evidence that the C II (1335) line also follows the radial velocity of the Si IV 1393 component.

The two emission features, C\,{\sc iv}  1548.202, (C\,{\sc iv}  1550.774 is very weak) reveal a clear shift as a function of orbital phase in the same {\it direction} as the motion of the white dwarf as seen in Fig.8.

Our study reveals that the emission lines of C IV (1548, 1550) vary significantly between STIS observations but their variation does not appear to depend upon orbital phase in any discernable way. However, at orbital phase 0.93, the C IV (1548, 1550) line went into absorption instead of the persistent emission. This 
transient absorption is likely associated with a blob of gas ejected by the K2V star and absorbing white dwarf photons as it passes across the line of sight of the white dwarf. The possiblity that a CME was detected is reinforced by the strong absorption at C IV (1548, 1550) since the CMEs give strong absorption at C IV. The STIS spectra reveal many more such possible CME events which will be discussed elsewhere (Mullan et al. 2012).

The measured radial velocities of the C\,{\sc iv}  1548 emission line are -26 km/s at orbital phase 0.227, -29 km/s at phase 0.268, 110 km/s at phase 0.692 and 81 km/s at phase 0.80. However, these velocities are much {\it smaller} than the WD velocity.  In Fig. 9, the He\,{\sc ii}  emission feature is displayed as a function of orbital phase. The He II emission is not strong enough to obtain reliable radial velocities but in the stacked spectra, it does line up with the observed shift of the absorption features like Si IV, which form in the photosphere.

Since the accreted metals darken the accretion regions in the X-ray and FUV spectral regions, their line strengths should be modulated on the X-ray rotation period of 555s, as a function of X-ray rotational phase. This rotational modulation of the line strength of accreted metals, specifically Si\,{\sc iii} 1206.5, was first seen in the HST GHRS spectra of V471 Tauri by Sion et al. 1998). In the STIS data which corresponds to X-ray rotational phase 0.0 (the primary accretion cap facing the observor), the Si\,{\sc iv} lines are deepest at phase 0.0566 
and phase 0.1715 indicating that the maximum line strength is slightly offset from 
rotational phase 0.0. The variation of the S\,{\sc iv} (1393, 1402) doublet as a function  of rotational phase are shown in Fig.10. The sharp components of 
Si\,{\sc iv} (1393, 1402) are broader than any interstellar features and, as stated earlier, are photospheric features. These sharp lines are blue shifted by 
$\sim 0.5-0.6$\AA\  and have widths of $\sim 0.35-0.5$\AA\ . 
On the other hand, the interstellar lines have a width of 
$\sim 0.1$\AA\  and a shift of $\sim 0.1$\AA\ . 

In Fig.11, the relatively weak 
Si\,{\sc  iii} (1206) feature is shown as a function of X-ray rotational phase
for phase 0.0 (WD magnetic pole pointed at the observor), 0.25, 0.50 and 0.75. 
The strongest absorption is seen at phase 0.0. In Fig.12, the same plot is given for 
C\,{\sc  ii} 1334, 1335 but note the contamination by the sharp interstellar features. Nonetheless, the absorption, while being weak, appears strongest at phase 0.0 In Fig.13 the C\,{\sc  iii} (1175.711) feature versus rotational phase reveals a strong absorption line at rotational phase 0.0. Note that there is no clearly detectable feature at phase 0.5 which corresponds to the secondary pole. 
In Fig.14, the variation of an Si III (1501) blend is shown versus rotational phase. At phase 0.0, the feature is quite strong. Although, aside from C III and Si IV, the absorption lines are too weak to reliably measure radial velocities, we note that all of the features in Figs.11, 12, 13, and 14 reveal the same modulation with rotational phase as the Si IV (1393, 1402) absorption lines,
suggesting the possiblity that they share the same origin in the photosphere of the magnetic white dwarf.

\subsection{Zeeman Splitting?}

It is curious that none of the absorption lines identified with the magnetized photosphere reveal clear Zeeman splitting. For example, the strongest absorption feature, the Si IV doublet, appears very broad and shallow with each component having a flat bottom $\sim 1.5$\AA\  in width. The Si\,{\sc iv} 1402\AA\ 
component appears asymmetric while the Si\,{\sc iv} 1393\AA\ 
component reveals no such asymmetry. The overall shape of the lines suggests some source of broadening, either rotational broadening, pressure broadening or a blend of Zeeman-split components. It is surprising that there is no clear evidence of Zeeman splitting given the earlier observational support for the origin of the 555s oscillations as being due to rotational modulation of magnetic accretion poles X-ray darkened by accreted metals (Robinson et al. 1988; Clemens 1992; Barstow et al. 1993; Sion et al. 1998). The Si\,{\sc iv} absorption features cannot be explained as simply due to rotational effects, since the width of the lines would require a rotational velocity of $\sim 250$ km/s while the 555 s X-ray rotational period of V471 Tau corresponds to a spin velocity of only about $80-100$ km/s for a 0.84
$M_{\odot}$ white dwarf.

A possible interpretation is that the width of these lines is caused
by the blending of Zeeman sub-components corresponding to the
2p$^6$3s-2p$^6$3p transitions of Si\,{\sc iv}.  In the low
magnetic field LS coupling regime, the wavelength displacement from
the central line position $\lambda_0$ (in \AA) is given by (Leone et
al. 2000)
\[
  \Delta \lambda = 4.67 \times 10^{-13} g_{\rm eff} \lambda_0^2 B
\]
where, $g_{\rm eff}$ is the effective Land{\'e} factor and $B$ is the
magnetic field strength in Gauss. The Si~{\footnotesize IV} transition
at $\lambda$1393.8 ($^2$S$_{1/2}$ - $^2$P$_{3/2}$) has $g_{\rm
  eff}=\pm 5/3, \pm 1$ ($\sigma$ components) and $g_{\rm eff}=\pm 1/3$
($\pi$ components), producing 6 components in total. The transition at
$\lambda$1402.8 ($^2$S$_{1/2}$ - $^2$P$_{1/2}$) has $g_{\rm eff}=\pm
4/3$ ($\sigma$ components) and $g_{\rm eff}= \pm 2/3$ ($\pi$
components), producing in this case four components. Hence, a field of
a few $10^5$ Gauss, together with rotational broadening and magnetic
field spread over the stellar surface could easily explain the width
and the flat bottom observed in these lines.

For a field of $\sim 9\times 10^5$ Gauss, the largest Zeeman shift of
the Si~{\footnotesize IV} transitions at $\lambda$1393.8 and
$\lambda$1402.8 is $\sim 10$\% of the term splitting. However, for
$B> 10^5$ Gauss the LS coupling regime may break down in favour of
the Paschen-Back regime. In this case, transitions at $\lambda$1393.8
and $\lambda$1402.8 will be Zeeman triplets.

\section{Synthetic Spectral Fit to the WD}

From the STIS data, first we re-determined the effective temperature and surface gravity of the white dwarf using the model atmosphere codes TLUSTY (Hubeny 1988) to compute the atmospheric structure and SYNSPEC (Hubeny, Lanz, \& Jeffrey 1994; Hubeny \& Lanz 1995) to construct the synthetic spectra. The details of our $\chi^2$ minimization fitting procedures can be found in Sion et al. (1995).
In Fig.15 we display the best fitting white dwarf model to the coadded 
STIS data with $T_{eff} = 34,100$K, $Log g = 8.2$ to 8.3. The equatorial rotation velocity of the WD corrresponding to a spin period of 555s, the X-ray period, is only 90 km/s so that for accretion onto magnetically confined accretion regions presumed to be at high latitudes, the rotation is only about 50 km/s.

Using these same model atmosphere codes, we determined the silicon and carbon abundances by fitting the detailed line profiles of  Si\,{\sc iv} and C\,{\sc iii} in order to determine the photospheric abundance of accreted silicon and carbon
in the polar accretion region/spot. We adopted a two-component model in which a pure hydrogen photosphere covers 60\% of the stellar surface
visible at phase zero and an accreted cap covers 40\% of the
hemisphere. These surface areas are taken from the EUV lightcurve
analysis of Dupuis et al. (1992) and the analysis of Barstow et al. (1993). The abundances of Si and C were varied over the range 0.001$<$(Si/H)/(Si/H)$_{\odot}<$0.5 while all of the other elements were fixed at very low abundances ($< 0.01$ solar and He$<$0.001).  
The Si abundance is Si = 0.015 ($\pm$0.005), and the C abundance is C = 0.0003 ($\pm$0.0002). Thus the C abundance is 50 times smaller that the Si abundance. Moreover, 
the rotational velocity, needed to fit the Si\,{\sc iv} lines and C\,{\sc iii} lines, is $v\sin{i} = 250$km/sec. 
is five times faster than the rotational velocity implied by the 555 s X-ray/EUV/optical oscillations. The silicon abundance profile fits are displayed in Fig.16. If we assume the expected rotational velocity of the white dwarf ($\sim$90 km/s at the equator), then it is not possible to match the width and shape of the Si\,{\sc iv} profiles with Si abundances of 0.1 to 1.0 (solar) and the profile fits worsen even more if lower Si abundances are used. Again, this implies that some broadening of the features must be responsible. Unless some pressure broadening is invoked, the implication is that the Zeemen effect is present but the substructure is not resolved.

Using the solar abundance of Asplund et al.(2005), we estimate that the relative
abundance of silicon with respect to hydrogen is $1.62\times 10^{-6}$
on the accretion spots, which is much larger than what is expected from
radiative levitation alone. In the context of diffusion theory, the
measured silicon abundance suggests that the silicon is originating
from an external source to the white dwarf. In view of the relatively
large surface gravity of V471 Tauri, one does not expect radiative
levitation to support a silicon abundance (by number relative to
hydrogen) much larger than $10^{-8}$ (Chayer, Fontaine, \& Wesemael,
1995). Being in a close binary system with an active K dwarf as a
companion, it is reasonable to expect that the hot white dwarf may
accrete a substantial fraction of the wind and perhaps material
ejected during flares and CMEs. Assuming that we can neglect radiative
levitation, which is a safe approximation as just alluded, we can
estimate the accretion rate needed to explain the rather large
abundance of silicon residing on the accretion spots. We also assume
that the effect of a probable magnetic field can be neglected as the
field strength is not sufficiently large to have a significant effect
on the diffusion coefficient. With these assumptions, we can apply the
gravitational settling formalism developed by Paquette et al. (1986)
and Dupuis et al. (1993) in the context of the accretion-diffusion
model for white dwarfs. We obtain that the diffusion velocity of
silicon in the atmosphere of V471 Tauri is about 0.1 cm/s and the
e-folding time for gravitational settling is about 4.3 days.  The
diffusion timescale is therefore very short in the atmospheric zone
and one expects that silicon would rapidly settle out of the atmosphere
if accretion was turned off. Assuming that a steady-state is reached,
one can apply equation 4 of Dupuis et al. (1993) describing the
steady-state condition for gravit
ational settling in the case of a
constant accretion rate. This theory was developed in the context of
cooler white dwarfs with superficial convective zones. Here, we
effectively assume that the atmosphere over the accretion spot has a
constant silicon abundance and replace $\Delta M_{cz}$ by $\Delta
M_{atm}$ (total mass in the atmosphere) in the equation. The silicon
accretion rate needed to explain this abundance level of silicon is
about $7.2\times 10^{-20}$ $M_{\odot}$/yr or $4.5\times 10^{6}$ g/s.
More detailed diffusion calculations that are beyond the scope of this
paper will be required to properly take into account the potential
effect of the magnetic field on the accretion flow and on the
diffusion in the atmosphere.

\section{Conclusions}

The archival time-resolved HST STIS spectra of the 
magnetic white dwarf in the Hyades eclipsing-spectroscopic, post-common 
envelope binary V471 Tauri has yielded a host of new insights into magnetically controlled 
accretion onto the white dwarf. The external source of the metal-laden atmosphere is
the K2V companion which led us to consider phenomena associated with surface activity on the K dwarf
such as CMEs, flares and wind. 

The rapid rotation of the K2V dwarf drives an extremely
high level of magnetic activity relative to the solar analogue (Walter 1994).
Indeed, the K2 V star has properties similar to those of the single
ultra-rapid rotators in young clusters, very young stars such as V410 Tauri (Skelly et al. 2010) and to the rapidly rotating K stars in RS CVn and BY Dra
systems.  Doppler images by Ramseyer et al (1995) reveal a large high-latitude spot. Ramseyer et al (1995) determined a radius for the K dwarf about 20\% greater than normal for a K2 dwarf the age of the Hyades. This finding was confirmed using HST GHRS spectra by O'Brien et al.(2001) and Bond et al. (2001) who also derived an accurate mass for the magnetically active, spotted K dwarf. 
The very fast rotation rate and depth of its convection zone are almost certainly the most important parameters in determining the strength and
configuration of the magnetic field on the rapidly rotating K dwarf (Skelly et al. 2010).

Walter (2001) explored the structure of the transition region/chromospheric gas with HST STIS. 
The HST STIS spectrum during the primary eclipse reveals a host of transition region/chromospheric
emission features including N V (1238, 1242), Si IV (1393, 1402), C IV (1548, 1550) and He II (1640) reported by Walter (2001, 2004). 
C IV (1548, 1550) and He II (1640). However, In addition, weaker emission lines are seen from cooler regions
such as C III (1175.71), Si III (1206.53), Si III (1304.37), O I (1304.86), C II (1334.53) and C II (1335.71).   
We find that the spectroscopic characteristics and emission line fluxes of the transition region/chromosphere 
of the K2V component of V471 Tauri are comparable to the emission characteristics of the 
fastest rotating K dwarfs in young open clusters.  

We have detected a number of absorption features associated with metals accreted onto the photosphere of the magnetic white dwarf. 
Among these features are C III (1175), Si (1206), C II (1335), Si IV (1393, 1402), and a feature at 1501 we identify as a blend of 
three Si III transitions. The C III (1175) and Si IV (1393, 1402) features are the strongest in the FUV spectrum and provide 
abundance determination. All other absorption lines are either interstellar or associated with a region above the white dwarf and/or 
with coronal mass ejection events illuminated as they pass in front of the white dwarf.  
We derived radial velocities for all of the absorption features due to accreted metals and for the emission lines due to C IV (1548, 1550) 
and He II (1640). The two emission lines reveal a clear shift as a function of orbital phase in the same {\it direction} as the motion of 
the white dwarf but with velocities that are much {\it smaller} than the WD velocity. All of the absorption features are modulated on the 
555s rotation period of the white dwarf with maximum line strength at rotational phase 0.0 when the primary magnetic accretion region is 
facing the observer. The maximum absorption strengths coincides with minimum light in the soft X-ray and EUV light curves of the WD and 
with maximum light in the optical light curve, thus providing further validity to the model in which the magnetic poles are darkened at 
short wavelengths by helium and metallic absorption and brightened in the optical by flux redistribution. 

It is puzzling that he photospheric absorption features show no clear evidence of Zeeman splitting.
The Si\,{\sc iv} and C\,{\sc iii} features are broad and flat-bottomed with no detectable Zeeman splitting. However, we point out that it is likely that a 
field of a few $10^5$ Gauss, together with rotational broadening and magnetic field spread over the stellar surface could easily explain the width, unresolved substructure and the flat bottoms observed in these lines. The absorption lines show no evidence of a correlation between their variations 
in strength and orbital phase. We report clear evidence of a secondary accretion pole and re-derive an effective temperature and spectroscopically-derived 
surface gravity for the magnetic white dwarf of $T_{eff} = 34,100$K +/- 100K, $Log g = 8.25$ +/- 0.05. We derive radial velocities and C and 
Si abundance from the Si\,{\sc iv} features. Our derived Si abundances of 0.015$\times$solar from the STIS spectra 
is seven times smaller than the Si abundance of 0.1$\times$solar derived by Sion et al. (1998) for the Si\,{\sc iii} (1206) feature in the GHRS spectra. This lower Si abundance together with the flux levels of the STIS observations indicate that V471 Tauri was less active during the STIS observations than the epoch when the GHRS spectra were obtained. 
If the accretion rate and diffusion rate of Si and C are in equlibrium, the accretion rate required to provide the C and Si is a measure of the efficiency of magnetic accretion and is four orders of magnitude smaller than the Bondi-Hoyle fluid rate, which implicates a magnetic-centrifugal propeller to greatly reduce the accretion efficiency. The accreted accreted ions provides crucial information on element diffusion in the presence of a magnetic field. 

This work was supported by HST Archival Grant HST-AR-11280.01-A and in part by NSF grant AST0807892 to Villanova University.

\section{References}

Asplund. M., Grevesse, N., Sauval, A.J. 2005, in : Cosmic abundances as records of stellar evolution and nucleosynthesis, Bash FN, Barnes TG (eds.), ASP Conf. series vol. 336, 25.

Ayres, T. R.; Linsky, J. L.; Simon, T.; Jordan, C.; Brown, A.1983, ApJ, 274, 784 

Ayres, Thomas R., Simon, Theodore, Stauffer, J. R., Stern, R. A., Pye, J. P., Brown, Alexander.1996, ApJ, 473, 279

Barstow, M.A. et al. 1992, MNRAS, 255, 369

Bond, H.E. 1985, in Cataclysmic Variables and Low-Mass X-Ray Binaries, ed. D.Q. Lamb
\& J. Patterson (Dordrecht: Reidel), p. 15

Bond, H.E., MUllan, D., O'Brien, S.\& Sion, E.M. 2001, ApJ, 560, 919

Chayer, P., Fontaine, G., \& Wesemael, F.1995, ApJS, 99, 189

Clemens, C. et al. 1992, ApJ, 391, 773

Dupuis, J., Vennes, S, \& Bowyer, S.1992, BAAS, 181, 1203

Dupuis, J., Fontaine, G., Wesemael, F.1993, ApJS, 87, 345 

Guinan, E.F., \& Sion, E.M. 1984, AJ, 89, 1252

Hric, L., Kundra,  E., \& Dubovsk´y, P.2011
Contrib. Astron. Obs. Skalnat´e Pleso 41, 39

Hubeny, I. 1988, Comput. Phys. Commun.,52,103

Hubeny, I., \& Lanz, T., \& Jeffrey, F. 1995, ApJ, 439, 875

Ibanoglu et al. 2005, MNRAS, 360, 1077

Iben, I., \& Livio, M. 1993, PASP, 105, 1373

Jeffries,R., Jackson, K., Briggs, P.,Evans,J. P. Pye3, 2010, MNRAS, 411, 2029.

Kim, J.S.,\& walter, F.M.1998, ASPC, 154, 1431

Lanning, H.H., \& Pesch, P. 1981, ApJ, 244, 280

Leone, F., Catanzaro, G., \& Catalano, S. 2000, A\&A, 355, 315

O'Brien, S. et al.2001, ApJ, 563, 971

Paczynski, B. 1976, in IAU Symp. 73, Structure and Evolution of Close Binary Systems, ed.P. Eggleton, S. Mitton, \& J. Whelan (Dordrecht:Reidel), p. 75

Paquette, C., Pelletier, C., Fontaine, G., \& Michaud, G.1986, ApJS, 61, 197

O'Brien, S., Bond, H., \& Sion, E.M.2001, ApJ, 563, 971

Pringle, J.E., \& Rees, M.J. 1972, A\&A, 21, 1

Robinson, R. D., Carpenter, K., Slee, O., Nelson,J., Stewart, R. 1994, MNRAS, 267, 918 

Schmidt, G. (+ 16 co-authors), 2005, ApJ, 630, 1037

Schreiber, M., \& Gaensicke, B.2003, A\&A, 406, 305

Simon, T. \& Fekel, F.1987, ApJ, 316, 434 
  
Sion et al. 1998, ApJL, 496, L29

Skelly, M., Donati, J.-F.,  Bouvier, J. Grankin, K., Unruh, Y., Artemenko, S., \& Petrov, P.2010, MNRAS, 403, 159 

Soderblom, D., Jones, B.F., \& Fischer, D.2001, ApJ, 563, 334

Walter, F.2001, in The Future of Cool-Star Astrophysics: 12th Cambridge Workshop on Cool Stars, Stellar Systems, and the Sun (2001 July 30 - August 3),    
    eds. A. Brown, G.M. Harper, and T.R. Ayres (University of Colorado), 2003, p. 14-39.

Walter, F.2004, AN, 325, 241

Wesemael, F., \& Truran, J.1982, ApJ, 262, L53

\section{Figure Captions}

\begin{figure}        
\plotone{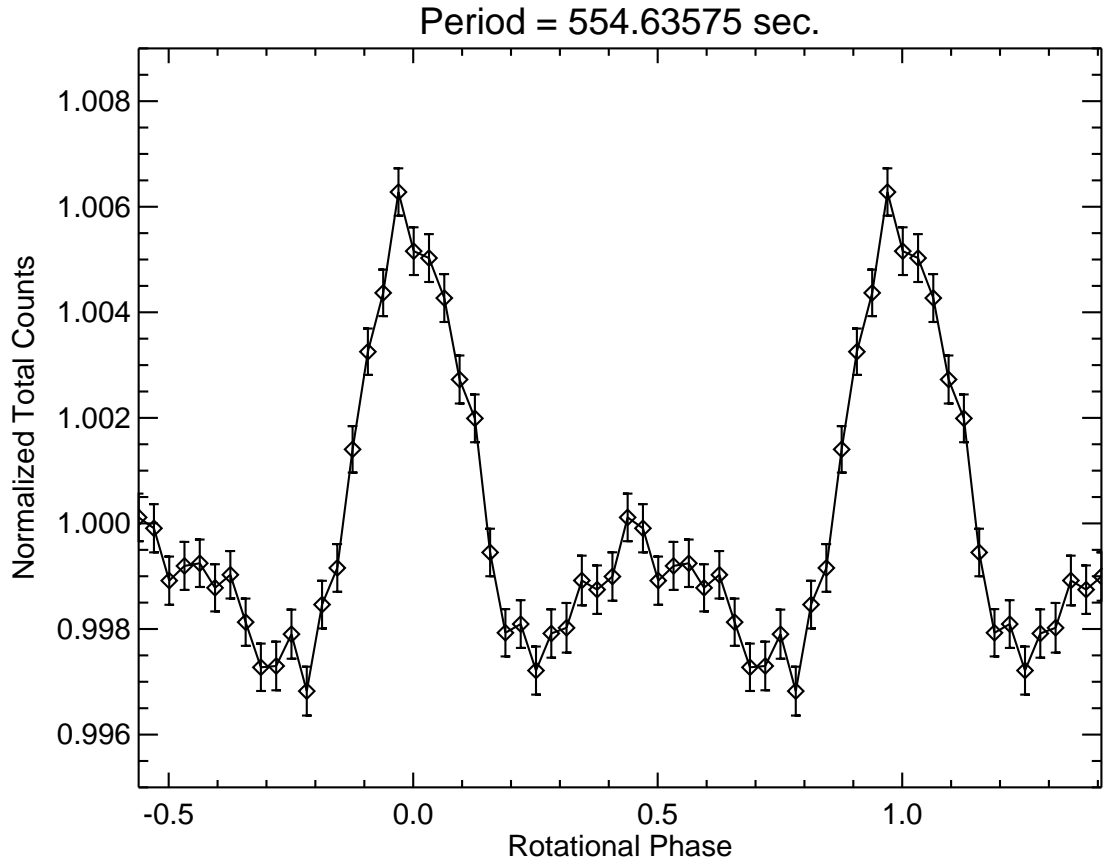}  
\caption{
Two phases of the rotational profile. showing the signature of the primary and secondary accretion poles.
Ephemeris: HJED 2450885.67019 + 0.0064193953E (computed from the STIS data)
}
\end{figure} 

\begin{figure}      
\plotone{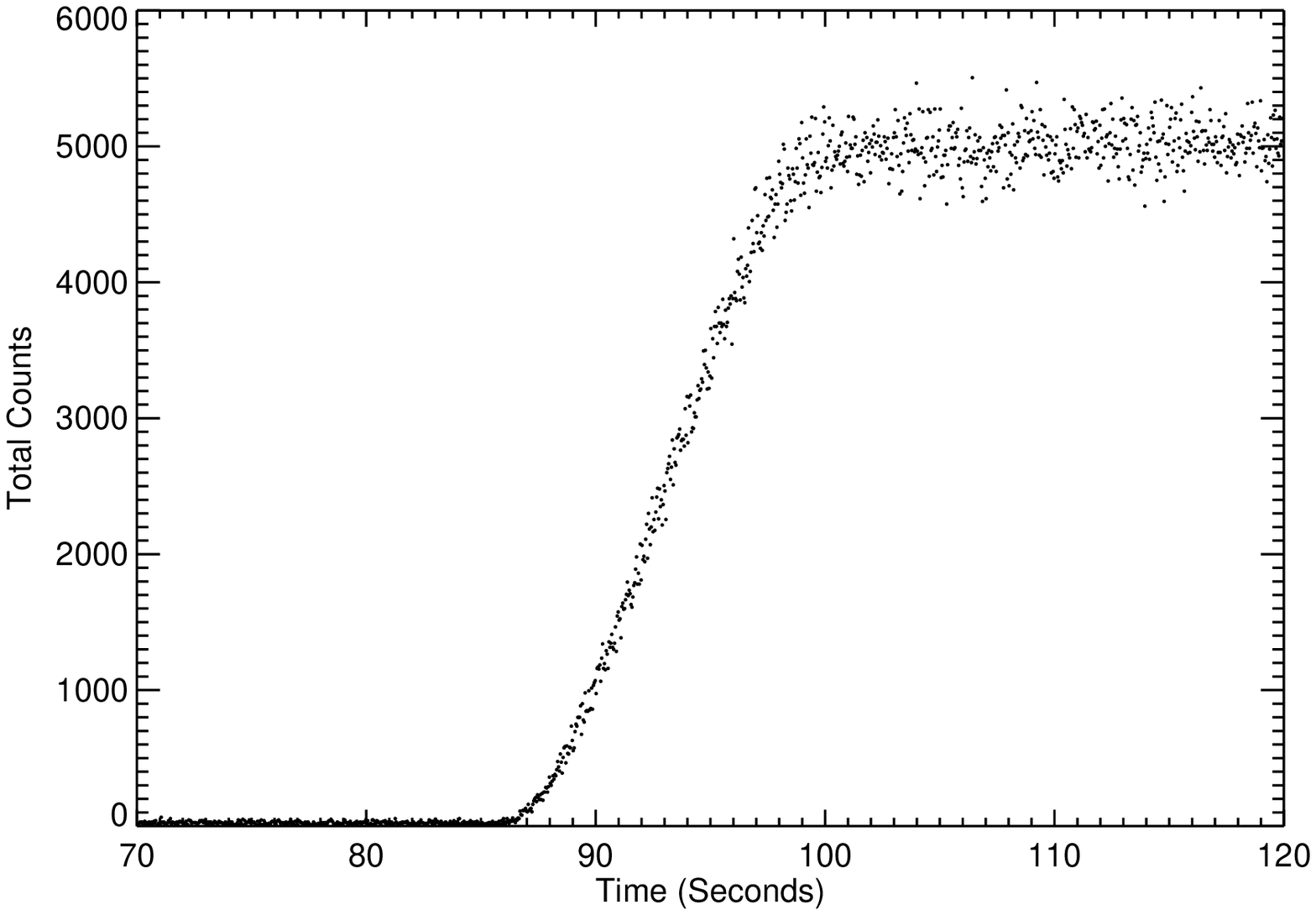}  
\caption{
Egress for eclipse (1998 data), number of counts versus time in seconds.
Data is tabulated on 0.2 second intervals.
}
\end{figure}

\begin{figure}       
\plotone{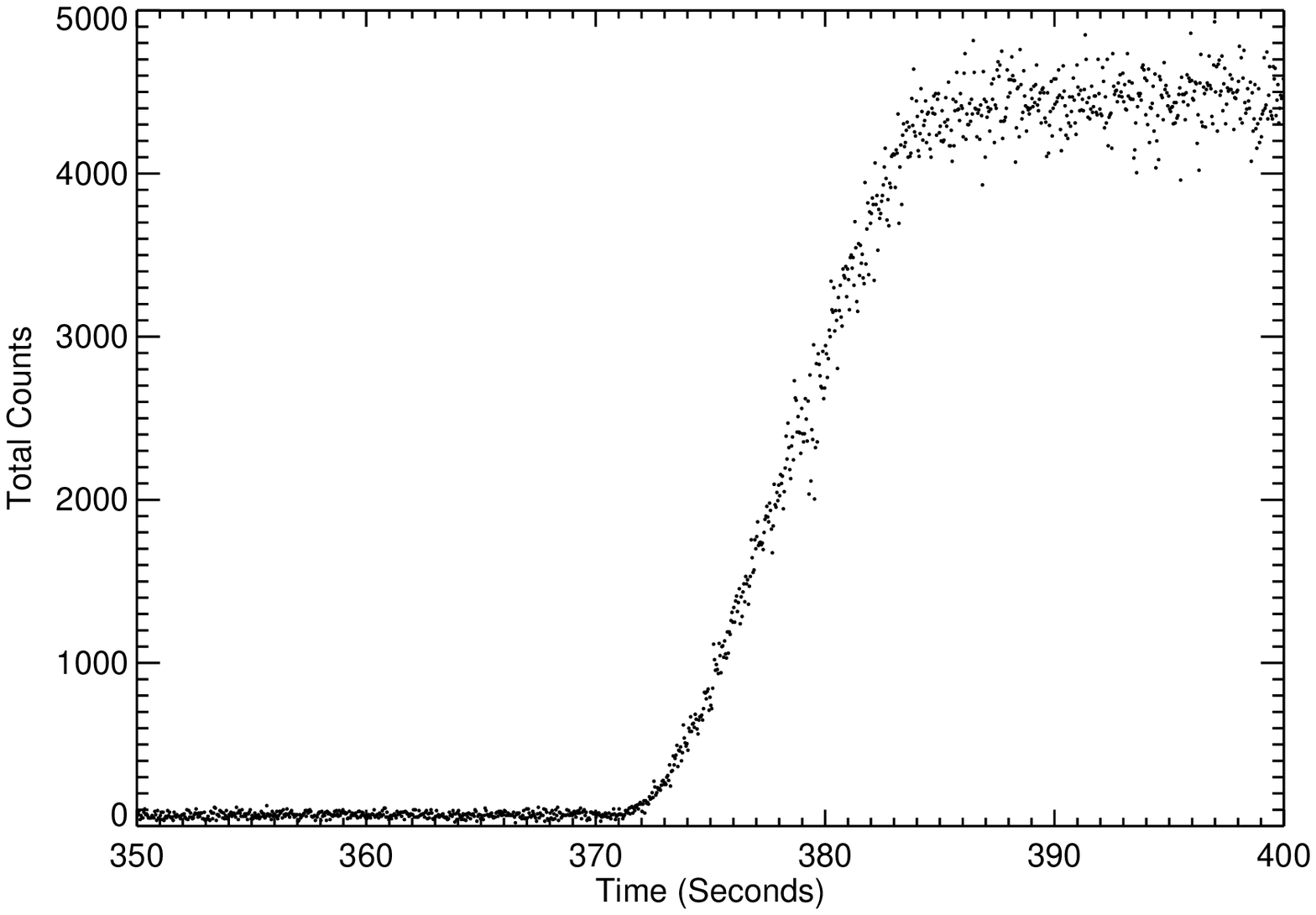} 
\caption{
Egress for eclipse (2002 data), number of counts versus time in seconds.
Data is tabulated on 0.2 second intervals.
}
\end{figure} 

\begin{figure}       
\plotone{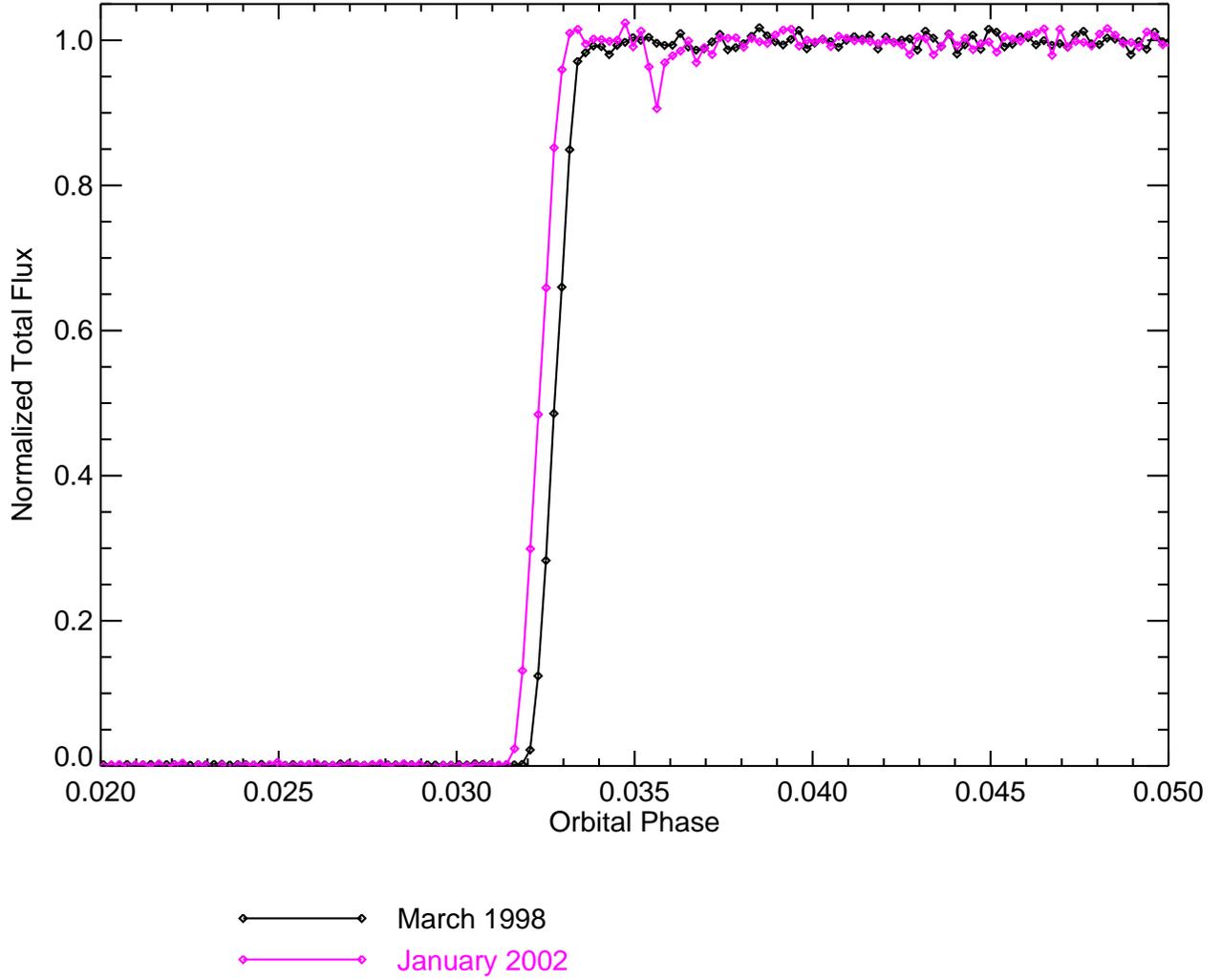}
\caption{
Eclipse egress versus orbital phase for the 1998 and 2002 data sets. An overplot is given of the 1998 
and 2002 eclipse data versus orbital phase. 
}
\end{figure}

\begin{figure}      
\plotone{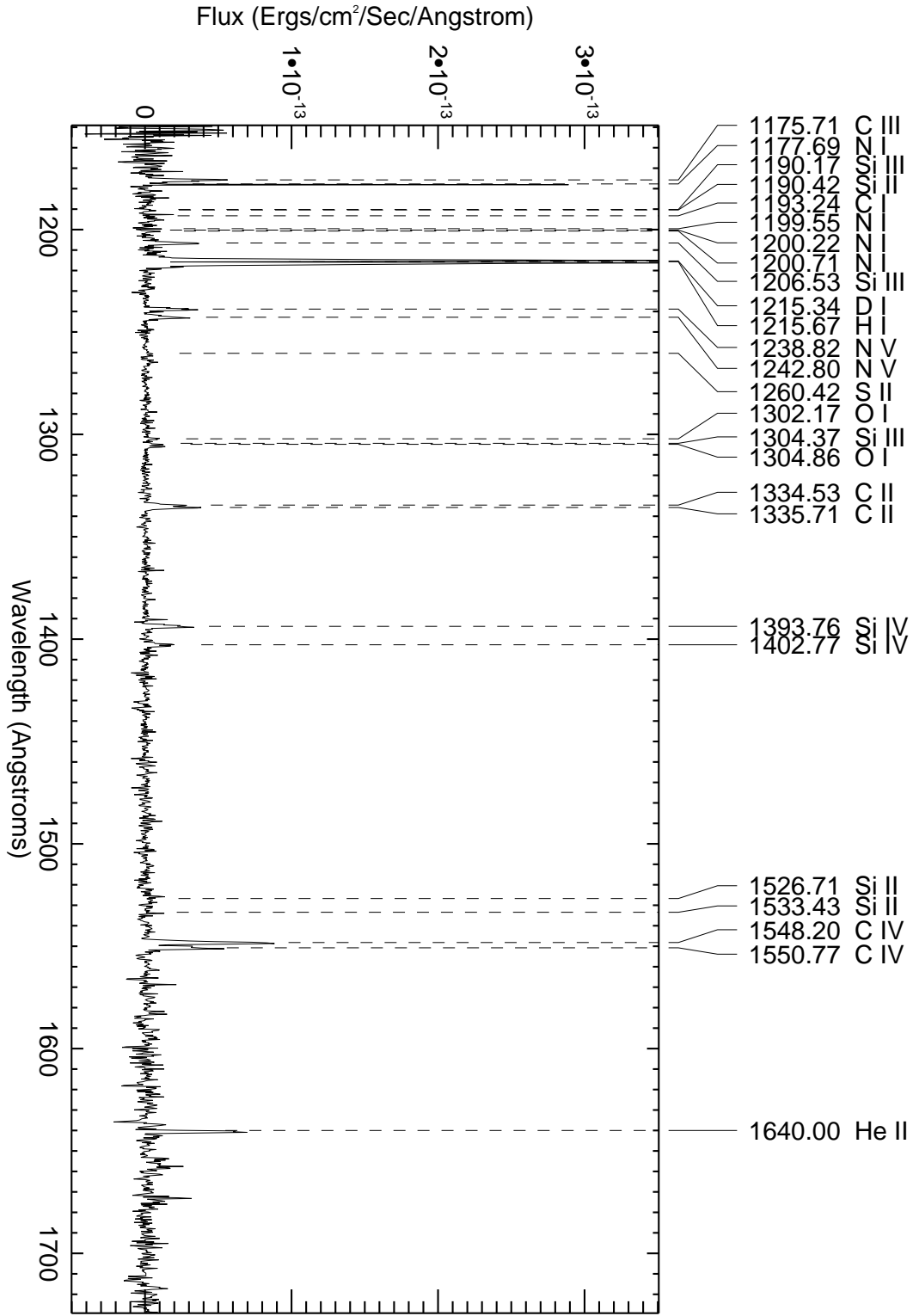}
\caption{The STIS spectrum obtained during the primary eclipse. The emission lines arise in the 
K2V star's chromosphere.
See text for details}
\end{figure}

\begin{figure}           
\plotone{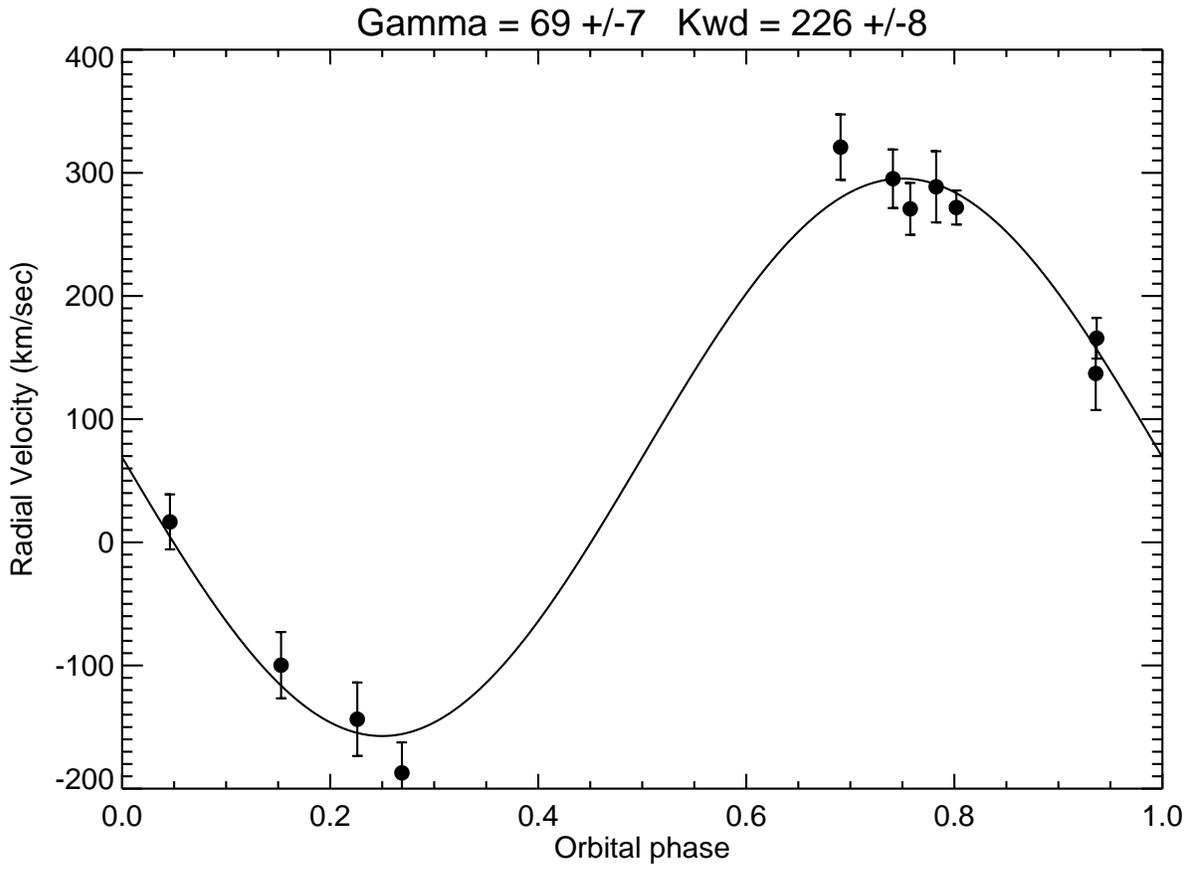}   
\caption{
Radial velocities of the SI\,{\sc iv} 1393.755 line fit with a sine wave.
}
\end{figure}

\begin{figure}           
\plotone{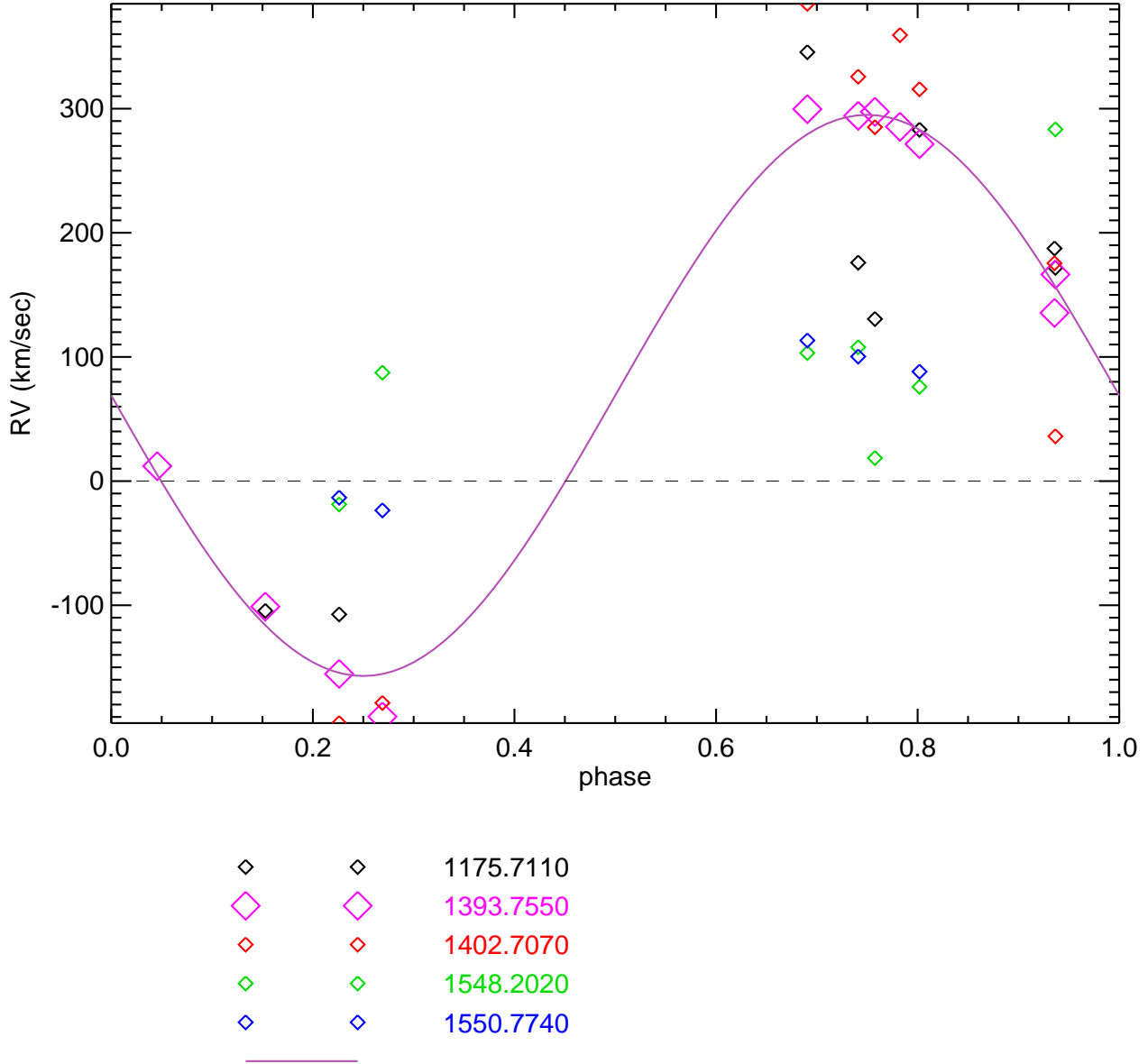}      
\caption{the radial velocities versus orbital phase for C III (1175), Si IV (1393, 1402), and C IV (1548, 1550) along with the 
radial velocity function computed using the SI IV 1393 line as shown in Fig.5b. The measurements were made using the data 
between rotational phases of 0.8 to 1.2 where the absorption is the strongest
}
\end{figure} 

\begin{figure}        
\plotone{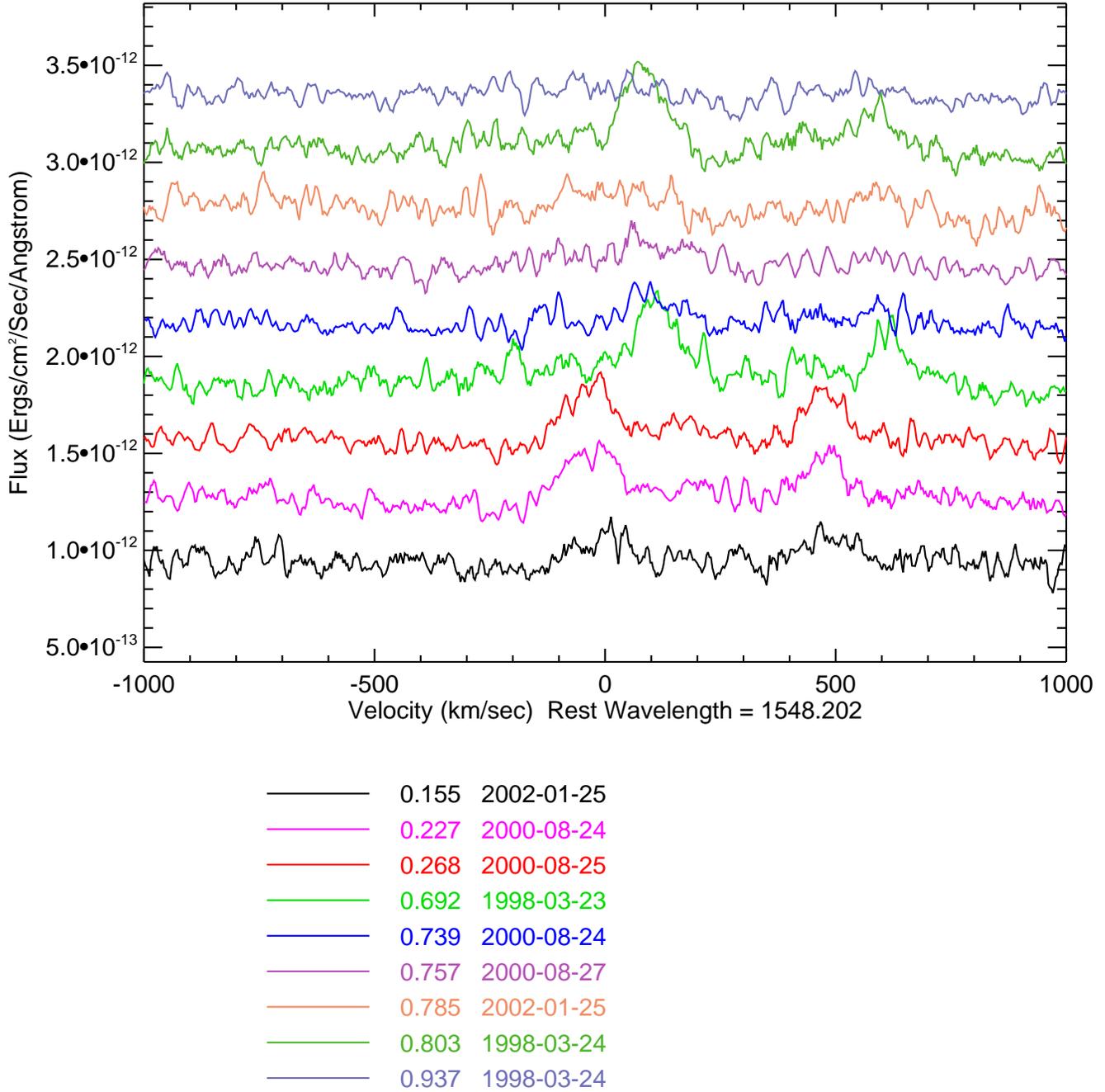}     
\caption{
  C\,{\sc iv}  1548.202 versus orbital phase (excluding CME observations). The 1550 component of 
the doublet is too weak to measure reliably.
}
\end{figure}

\begin{figure}       
\plotone{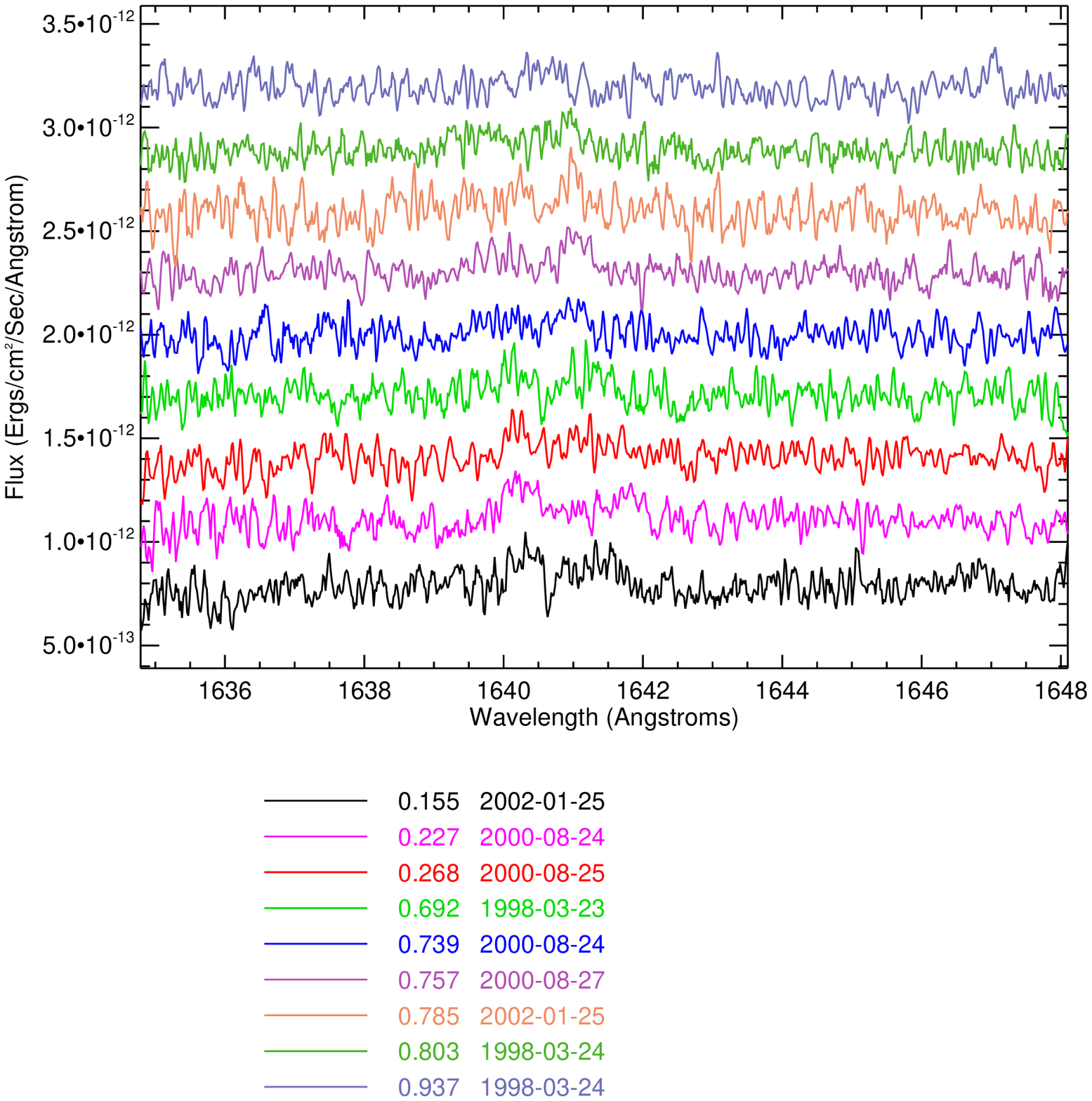}    
\caption{
 the He\,{\sc ii} 1640 emission  versus orbital phase. The feature is not strong enough to obtain accurate radial velocities.
}
\end{figure}

\begin{figure}           
\plotone{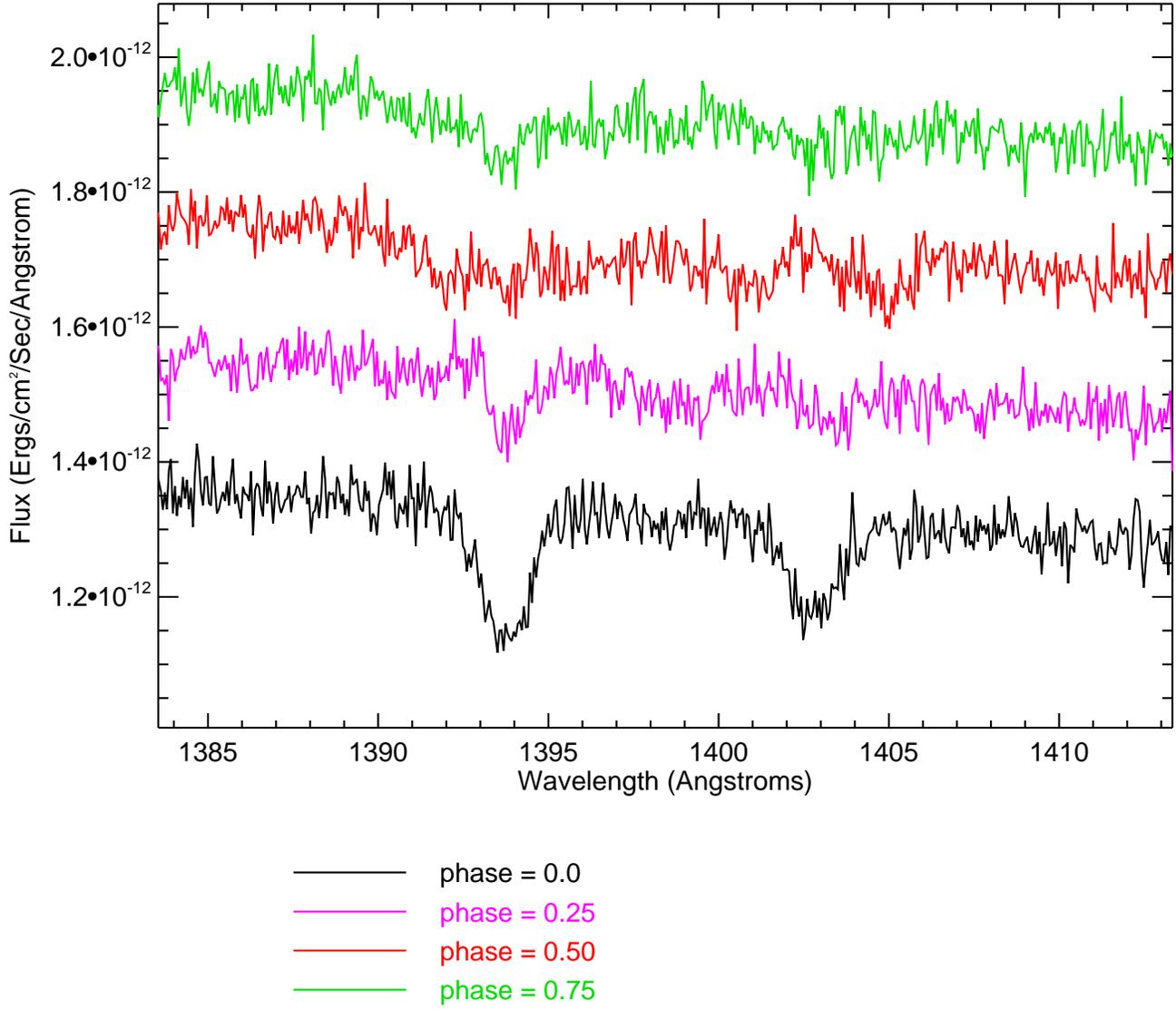}   
\caption{The variation of the S iv (1393, 1402) doublet as a
function of X-ray rotational phase.}
\end{figure} 

\begin{figure}        
\plotone{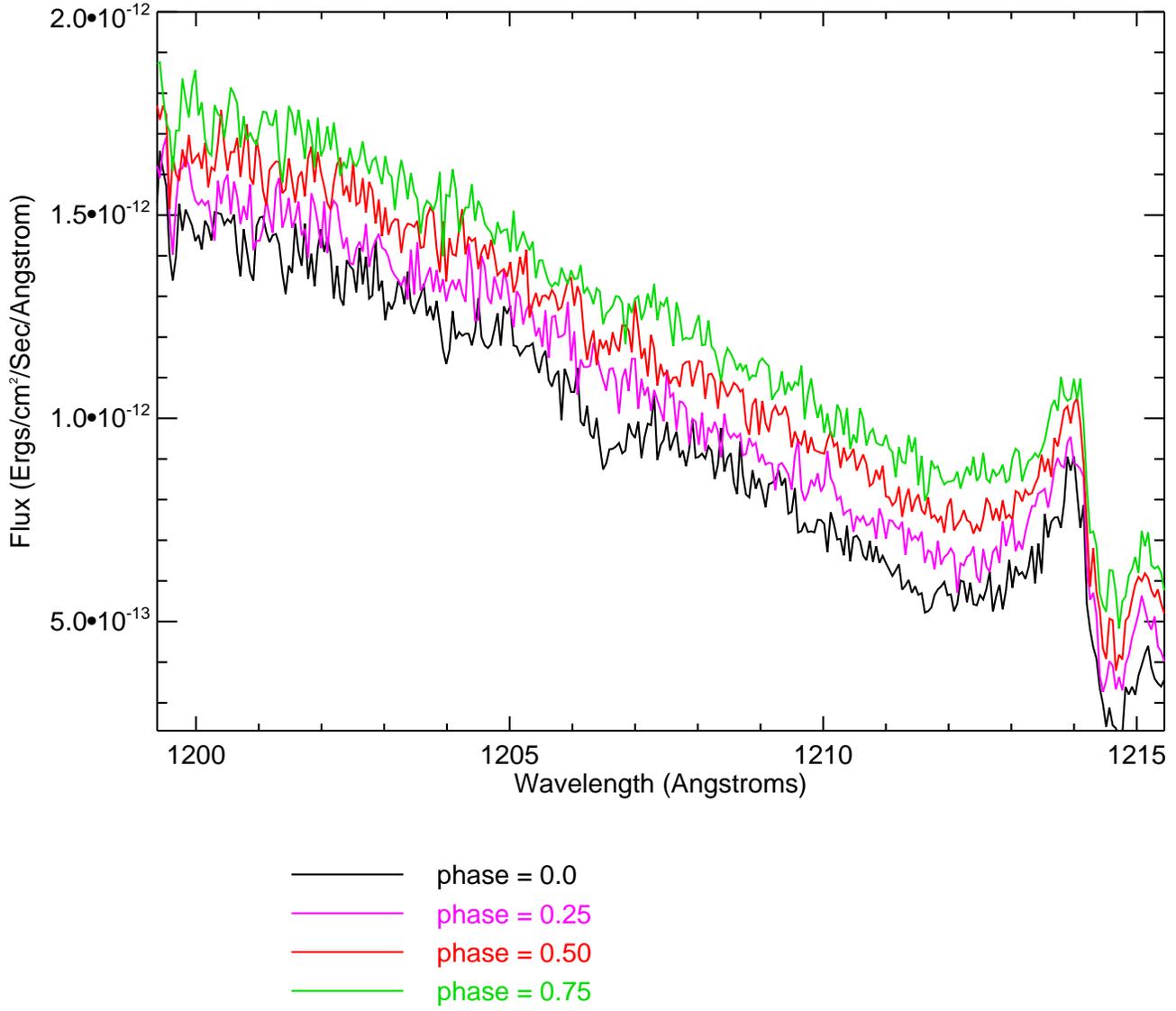}   
\caption{
 Si\,{\sc iii} 1206 versus rotational phase with the absorption strongest at rotation phase 0.0.
}
\end{figure} 

\begin{figure}        
\plotone{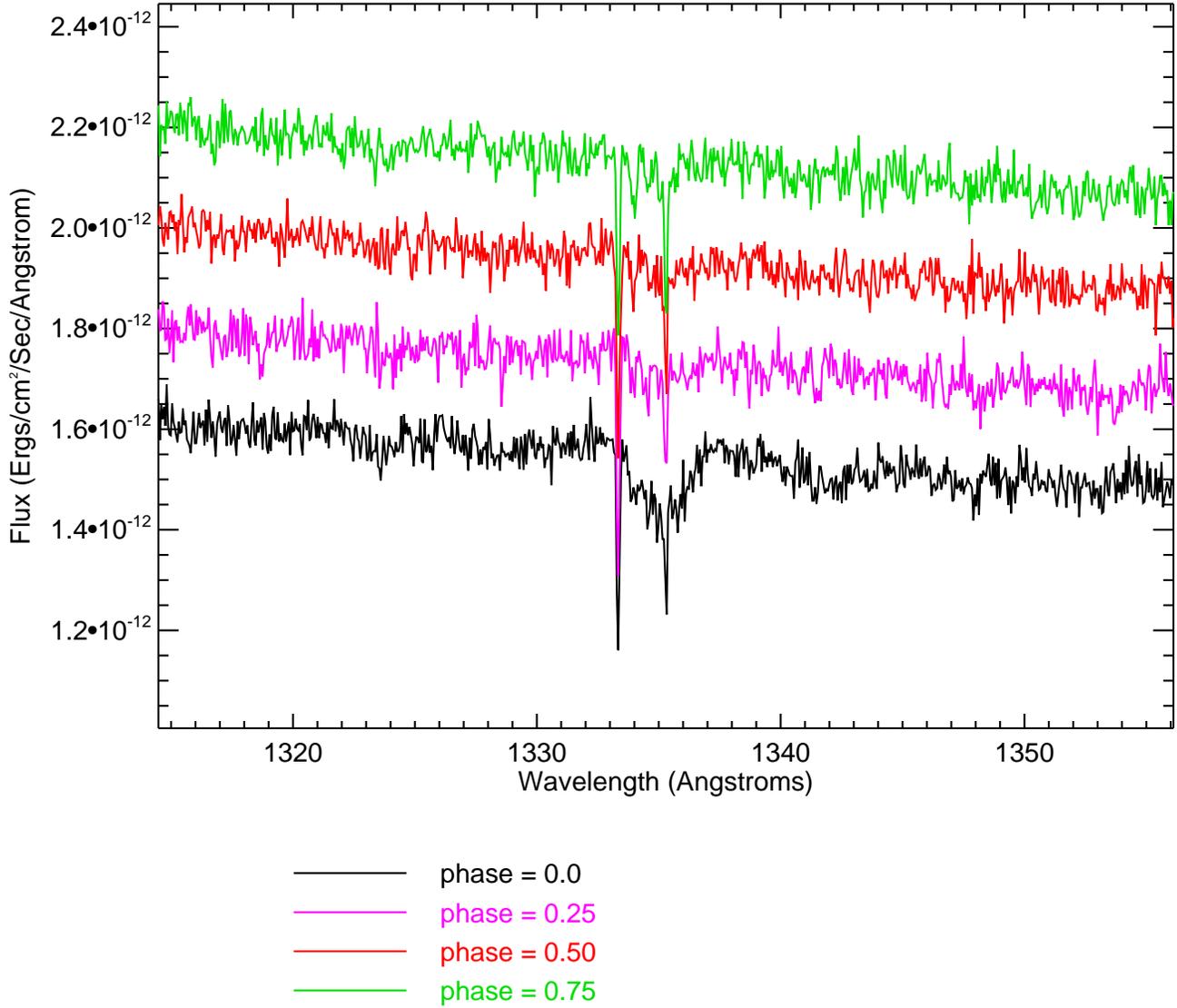}  
\caption{
 C\,{\sc ii} 1334.5323,1335.7077 versus rotational phase. Note the presence of very sharp, strong interstellar lines.
}
\end{figure} 

\begin{figure}        
\plotone{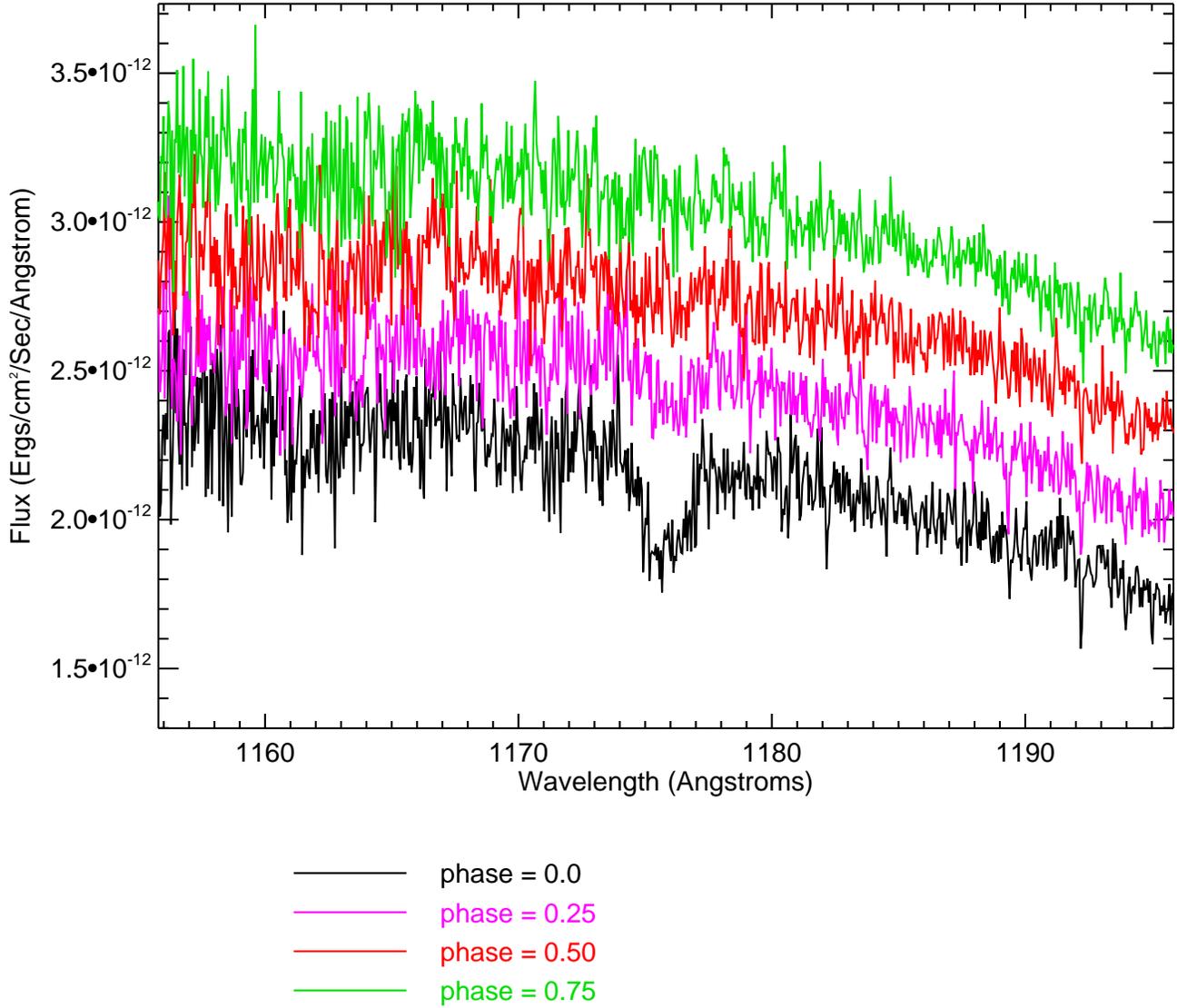}    
\caption{
 C\,{\sc iii} 1175.711 versus rotational phase with maximum strength at rotation phase 0.0. This feature is the strongest line absorption
in the STIS spectrum.
}
\end{figure} 

\begin{figure}        
\plotone{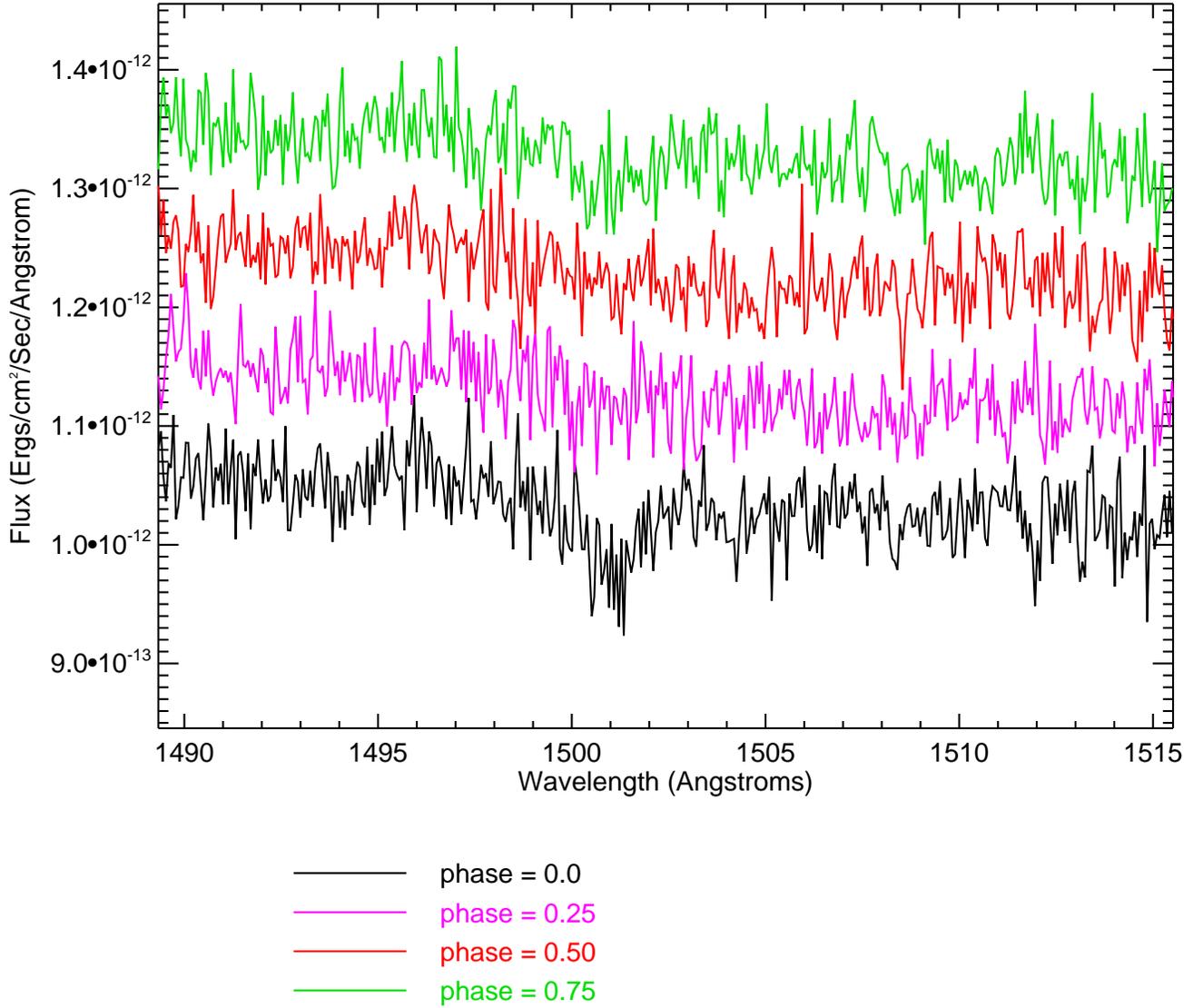}     
\caption{
 Si III Line at 1501\AA\  versus rotational phase. The identification of Si III is tentative since P III has two strong transitions
at 1501A as well.
}
\end{figure}

\begin{figure}         
\plotone{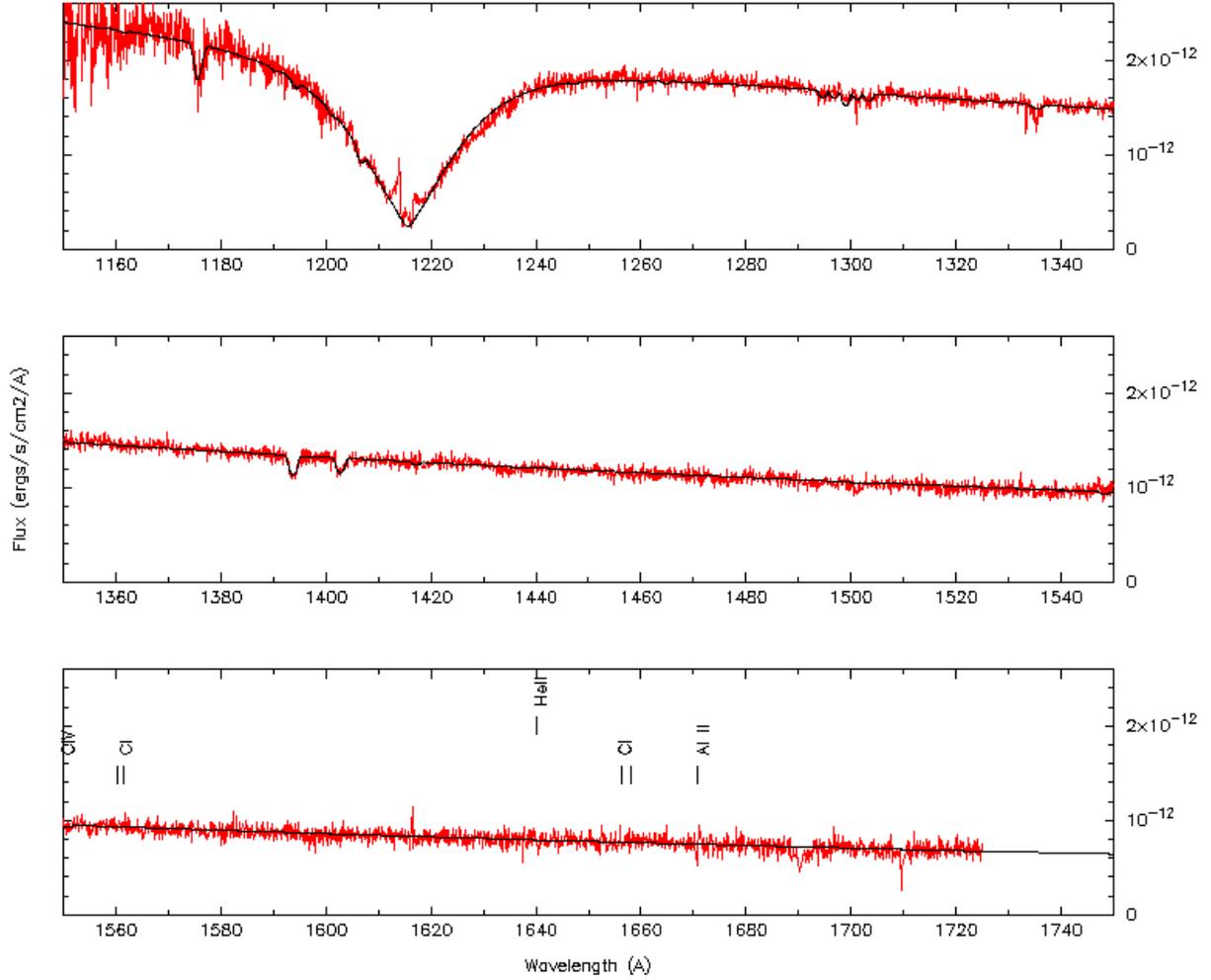}  
\caption{The best-fitting white dwarf synthetic spectral model (
$T_{eff} = 34,100$K, $Log g = 8.25$) to the full wavelength range of the STIS spectrum (see text for details)}
\end{figure} 

\begin{figure}          
\plotone{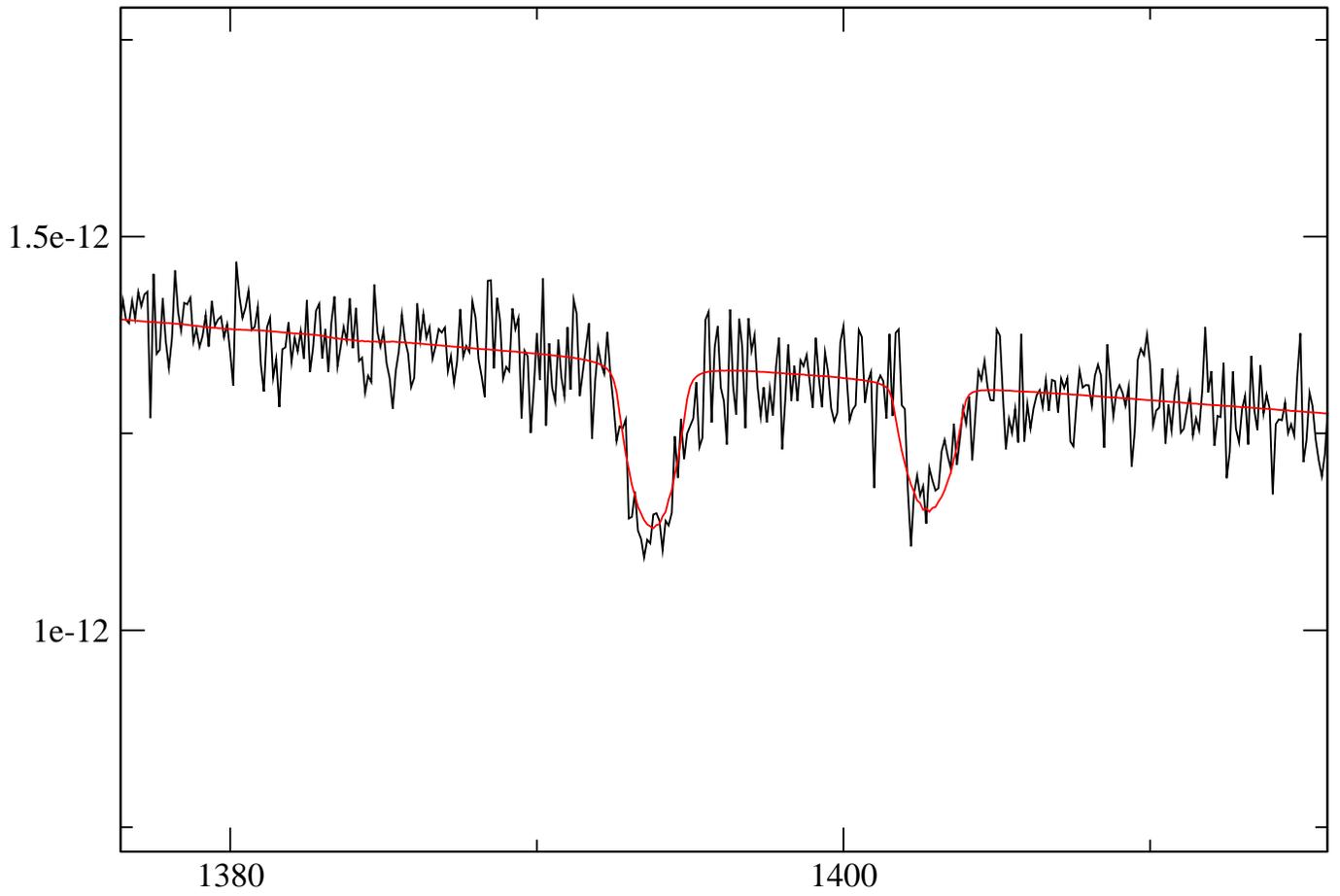}  
\caption{
The best-fitting model fluxes to the Si\,{\sc iv}  line profiles in the STIS spectrum (see text for details).
}
\end{figure} 

\end{document}